\documentclass[12pt,superscriptaddress,showpacs]{iopart}
\usepackage{graphicx,iopams}
\usepackage{bm}
\input epsf.sty
\begin{document}

\title{Possible $s^{\pm}$-wave pairing evidenced by midgap surface bound states in Fe-pnictide superconductors}
\author{C. S. Liu}
\address{College of Science, Yanshan University, Qinhuangdao 066004, China}
\author{J. Y. Chang}
\address{Department of Physics, Yanshan University, Qinhuangdao 066004, China}
\author{W. C. Wu}
\address{Department of Physics, National Taiwan Normal University, Taipei 11677, Taiwan}
\author{Chung-Yu Mou}
\address{Department of Physics, National Tsing Hua University, Hsinchu 30013,
Taiwan}\address{Institute of Physics, Academia Sinica, Nankang 11529, Taiwan} \address{Physics
Division, National Center for Theoretical Sciences, P.O.Box 2-131, Hsinchu, Taiwan}

\begin{abstract}
A phenomenological theory of tunneling spectroscopy 
for Fe-pnictide superconductors is developed by taking into consideration of asymmetric interface scattering
between particle and holes. It is shown that in consistent with anti-phase $s^{\pm}$-wave pairing, appreciable zero-energy surface bound states exist on the [100] surface of Fe-pnictide superconductors.
However,  in contrast to the [110] bound states in $d$-wave cuprate superconductors,
these bound states arise as a result of non-conservation of momentum
perpendicular to the interface for tunneling electrons and the
$s^{\pm}$ pairing, and hence they can only exist in a small window
($\sim \pm 6^{\circ}$) in the orientation of edges near [100]
direction. Our results explain why zero-bias conductance peak is
often observed in tunneling spectroscopy and when it disappears, two
coherent peaks show up. These results provide unambiguous signals to test the
possible $s^{\pm}$-wave pairing in Fe-pnictide superconductors.
\end{abstract}

\maketitle

\section{Introduction}

High temperature superconductivity has been recently observed in several classes of
Fe-pnictide materials\cite{RMP}. One of the key issues
towards understanding the superconductivity in these systems lies in identifying the pairing symmetry of the Cooper pairs. However, up to now, gap symmetries obtained from experimental observations show remarkable dependence on material classes and doping levels\cite{hicks-2009,
PhysRevLett.101.047003, hashimoto:017002} and a conclusive determination of the pairing
symmetry remains unsettled. Among various candidates,
the most natural and promising pairing state is considered to be the
$s^{\pm}$-wave in which the superconducting (SC) gap exhibits a 
sign reversal between $\alpha$ and
$\beta$ bands and can be naturally explained by the spin fluctuation
mechanism\cite{1367-2630-11-2-025009, PhysRevLett.101.057003, wang:047005,
PhysRevB.81.184512, Tsuei2010}.

Experimentally, point-contact Andreev-reflection spectroscopy (PCARS) is considered as
one of the high-resolution phase-sensitive probes for detecting the SC pairing state.
For instance, zero-bias conductance peak (ZBCP) associated with the Andreev bound
states (ABS) has given a direct evidence on the
$d$-wave pairing of high-$T_c$ cuprate superconductors
\cite{Hu1526, Tanaka3451, RevModPhys.77.109, PhysRevB.66.012512, PhysRevB.67.024503}.
However, PCARS measurements have not
yielded consistent results on Fe-based superconductors. While some PCARS measurements showed two coherent peaks and indicate
that SC pairing state might be fully gaped on the Fermi surface (FS)\cite{
PhysRevB.79.012503, PhysRevLett.105.237002,springerlink:10.1007/s10948-009-0469-6},
there are also measurements showing the existence of ZBCP and implying
the presence of zero-energy bound
states or ABS on the interface\cite{0295-5075-83-5-57004, 0953-2048-21-9-092003, 0953-2048-23-5-054009}.
More intriguingly, depending on the direction of the sample interface,
some PCARS measurements exhibit the coexistence of ZBCP with finite-energy coherent
peaks\cite{0953-2048-22-1-015018, 1367-2630-11-2-025015}.

On the theoretical side, no consensus has yet been reached to understand
the PCARS data either.
Although it is commonly believed that surface bound states (midgap states) are responsible for the complex PCARS,
so far most theoretical
studies favor that, in contrast to $d$-wave ABS of zero energy,
surface bound states have finite energies for iron pnictides \cite{PhysRevLett.102.157002}.
Furthermore, these surface bound states are generally resulted from the inter-band coupling
that , the ways of coupling are complicated
in the anti-phase $s^{\pm}$-wave pairing system.

Among many theoretical works,
the study in Ref. \cite{PhysRevB.79.174529} believes the coupling are from the boundary.
The matching condition for the wave function at the interface is used to produce two-band coupling on the basis of an extension of quantum waveguide theory.
Based on the Green's function formulation in this study, the differential conductance curves versus bias voltage explicitly
support the emergence of ABS as a manifestation of interference effects
between the bands.
The other study, however, suggests that the coupling is due to the direct coupling of two-orbital coupling\cite{PhysRevB.79.174526}. In these models,
ABS was also found to
appear at surface due to the sign change in the gap function when taking the inter-band
quasi-particle (QP) scattering into account.
Inspired by the tunneling magnetoresistance where the
leads are usually transition metals with multi-band $d$ orbitals, a new type of two-band coupling is set up, in which it is assumed that when an electron crosses the interface from the lead to the Fe-based superconductors, it tunnels into the first or second band on the right with the ratio of probability amplitudes
$\alpha_0$\cite{PhysRevLett.103.077003}. It was found that 
Andreev bound states can appear at both nonzero and near zero energies
by changing values of $\alpha_0$.

To address these puzzling issues and account for the
observed ZBCP, in this paper, we 
propose a different coupling scheme of two orbitals in the $s^{\pm}$-wave pairing state 
by considering asymmetric interface scatterings between particles and holes.
 Theoretically, the interface scatterings
for quasi-particles and quasi-holes of the same band are generally different due to that there is no particle-hole symmetry
in the normal metal side. Asymmetry in the probability amplitude of two orbitals would induce asymmetry between particles and holes in the same orbital.   
By assuming asymmetry between particles and holes in each orbital, we extend the Blonder-Tinkham-Klapwijk (BTK) formalism \cite{PhysRevB.25.4515}
to investigate the differential conductance of the junction between a normal
metal and a Fe-pnictide superconductor.
It is shown that 
by including directional dependence of QPs interplaying between different
bands, ZBCP emerges in the presence of asymmetry between particles and holes. The existence of the surface bound state is due to the anti-phase $s^{\pm}$-wave pairing potentials of different bands and the orientation in terms of multi-band
FS topology. 
Our results are consistent with recent PCARS measurements (with zero energy or nonzero energy) in iron-pnictide
superconductors. In particular, it is shown that ZBCP is
sensitive to surface orientation.
Off the [100] direction, the zero-bias
peak disappears and is replaced by two coherent peaks. These features provide unambiguous signals to test the
possible $s^{\pm}$-wave pairing in Fe-pnictide superconductors.

The paper is organized as follows. In Sec.~\ref{Model and Method}, we present the
model and basic formalism for studying the tunneling conditions. This formalism is
based on the WKBJ approximation of the Bogoliubov-de Gennes (BdG) equations and
consider QPs interplaying between different bands. In Sec.~\ref{Results and Discussions}, we fist give the numerical solutions in Sec.~\ref{Numerical simulations} and then illustrate in 
Sec.~\ref{Zero energy bound state} the ZBCP existing due to a close Saint-James cycle in the $s^\pm$ pairing of two bands. At last in Sec.~\ref{Comparing with the experimental data}, conductance spectrum were computed and compared to the
existing experimental data. In particular,
we show how the ZBCP
can exist in a small window ($\sim \pm 6^{\circ}$) in the orientation of
edges near [100] direction and how the coherent peaks can be observed in nonzero-energy in some case.
A brief summary is given in Sec.~\ref{summary}.

\section{Model and Method} \label{Model and Method}

In contrast to cuprates whose low-energy electronic structure is dominated
by Cu $3d_{x^2-y^2}$ orbital, the electronic structure of Fe-based compounds involves
all five Fe $3d$ orbitals forming multiple FS sheets. The FS topology and gap opening are
observed to be slightly different between different material classes and compositions.
Taking the newly found Tl$_{0.58}$Rb$_{0.42}$Fe$_{1.72}$Se$_2$ and K$_{0.8}$Fe$_{1.7}$Se$_2$
samples for example, high-resolution ARPES measurements have found that FS
for Tl$_{0.58}$Rb$_{0.42}$Fe$_{1.72}$Se$_2$ consists of two electron-like FS sheets
around the $\Gamma$ point\cite{PhysRevLett.106.107001}. The FS
around the M point shows a nearly isotropic SC gap of $\sim$ 12 meV. The large FS
near the $\Gamma$ point also shows a nearly isotropic SC gap of $\sim$ 15 meV,
while there is no clear SC gap opening for the inner tiny FS. 
On the other hand, for K$_{0.8}$Fe$_{1.7}$Se$_2$, a nearly circular FS is formed around M point\cite{PhysRevLett.106.187001}.
The absence of a hole-like FS is because hole-like band is shifted down below the Fermi energy.
The FS and SC gap properties of Tl$_{0.58}$Rb$_{0.42}$Fe$_{1.72}$Se$_2$ and
K$_{0.8}$Fe$_{1.7}$Se$_2$ are quite different from those of the earlier found sample
Ba$_x$K$_{1-x}$Fe$_2$As$_2$, including all under-, optimally- and over-doped regimes
\cite{HDing09, PhysRevB.79.054517, singh:237003, 0295-5075-83-4-47001, PhysRevB.83.020501}.
For Ba$_x$K$_{1-x}$Fe$_2$As$_2$, two large circular FSs are found on the $\Gamma$
point and one little FS on the M point. The SC gap on each FS is nearly
isotropic and the gap value on each FS nearly scales with T$_c$ over a wide doping
range ($0.25 \leq x \leq 0.7$).

\begin{figure}[ptb]
\begin{center}
\includegraphics[width=9cm ]{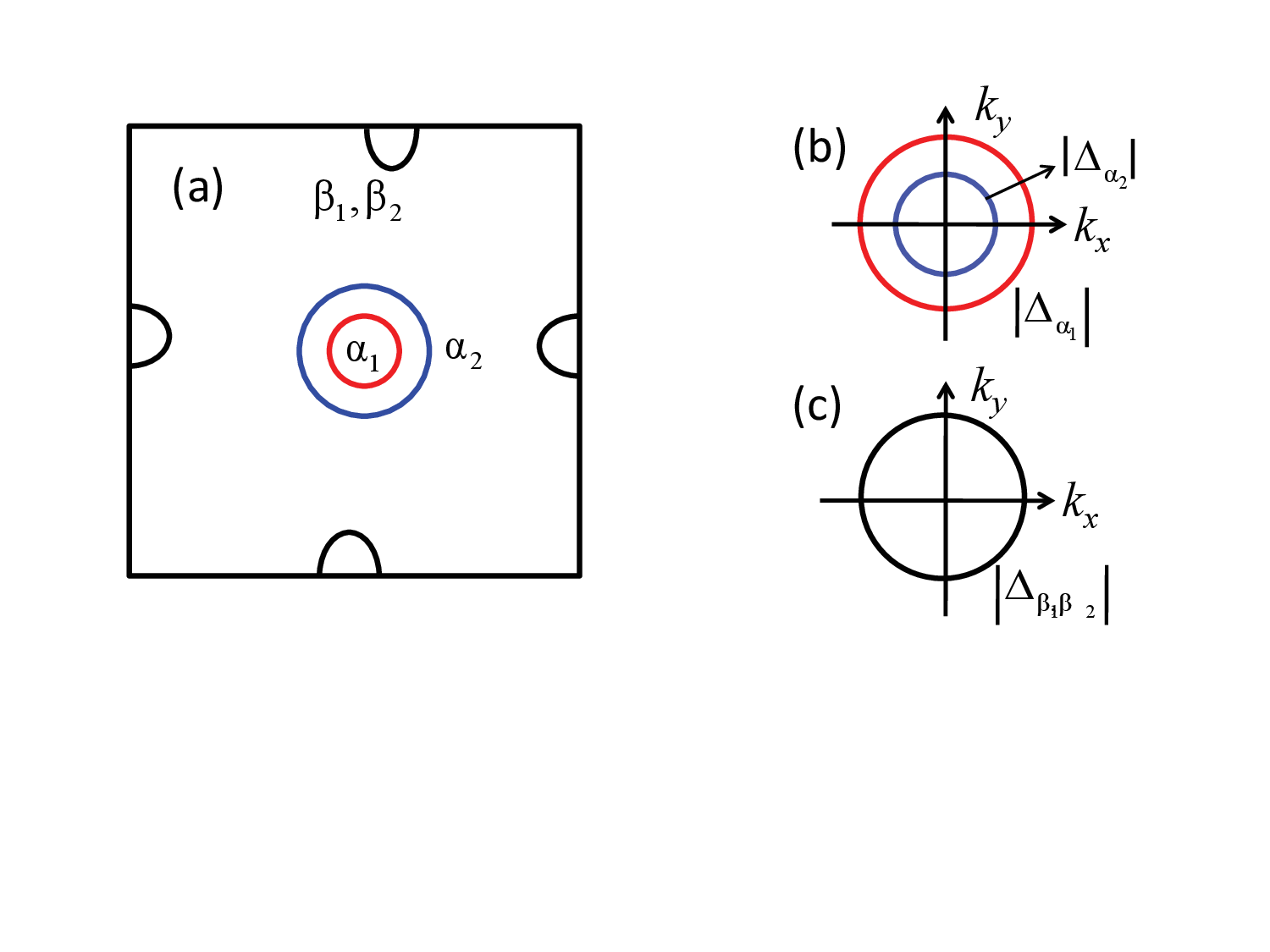}
\vspace{-2cm}
\caption{(Color online) Panel (a): schematic plot of the Fermi surfaces of two band in unfolded Brillouin zone. Panel (b) \& (c) plot the pairing amplitudes for $\alpha$ and $\beta$ band respectively.}
\label{fig1}
\end{center}
\end{figure}

To simplify discussions and capture the key elements, we use the following
model for Fe-pnictide superconductors.
The so-called $\alpha_1$ and $\alpha_2$ Fermi sheets are concentric and nearly
circular hole pockets around the $\Gamma$ point. While $\beta_1$ and $\beta_2$ Fermi
sheets are nearly circular electron pockets around the M points. SC gap is isotropic
on each FS but with different amplitude. For $s^{\pm}$-wave symmetry, gaps on
$\alpha$ and $\beta$ bands have a sign reversal and the corresponding magnitudes 
will be denoted by $\Delta_{\alpha}$ and $\Delta_{\beta}$.
Our discussions will follow the above picture
of FSs and SC gap structures, as sketched in Fig.~\ref{fig1}.
As it will be shown, slight distortion of the FS will not affect much the results.

We first consider a Fe-pnictide bulk superconductor with a perfectly
flat and infinitely large interface located at $x=0$. The superconductor occupies $x>0$. 
For {\em each} band, QP states have a coupled electron-hole character and can be
described by the BdG equations \cite{Gennes}
\begin{equation}
\left[
\begin{array}{cc}
\hat{\xi}  & \Delta \left(\mathbf{r}\right)  \\
\Delta^{\ast}\left(\mathbf{r}\right)  & -\hat{\xi}
\end{array}
\right] \left[
\begin{array}{c}
u\left( \mathbf{r}\right)  \\
v\left( \mathbf{r}\right)
\end{array}
\right] =E\left[
\begin{array}{c}
u\left( \mathbf{r}\right)  \\
v\left( \mathbf{r}\right)
\end{array}
\right],   \label{Bogoliubov-de Gennes equations}
\end{equation}
where $\Delta = \Delta_{\alpha}$ or $\Delta_{\beta}$, $E$ is the total energy of the quasi-particle, $\hat{\xi}\equiv -\hbar ^{2}\nabla^{2}/2m-\mu$ with $\mu$ being the chemical potential
potential, and $m$ the electron mass. Within the WKBJ approximation\cite{PhysRev.187.556},
one seeks solutions of the form
\begin{equation}
u\left( \mathbf{r}\right) =e^{i\mathbf{k}_{F}\cdot \mathbf{r}
}\eta \left( \mathbf{r}\right) {\rm ~~and~~}v\left( \mathbf{r}
\right) =e^{i\mathbf{k}_{F}\cdot \mathbf{r}}\chi \left( \mathbf{r}\right),
\end{equation}
where $\mathbf{k}_{F}$ is the Fermi wavevector satisfying 
$\mu=\hbar^2 k^2_F /(2m) $ and
in contrast to the plane-wave exponential factors,
$\eta \left( \mathbf{r}\right)$ and $\chi \left( \mathbf{r}\right)$ are slowly
varying functions. We shall neglect the difference of $k_F$ in  $\alpha$ and $\beta$ bands and the second derivatives in Eq. (\ref{Bogoliubov-de Gennes equations}). We then obtain the Andreev equations
\begin{eqnarray}
\left[ i\hbar \left( \mathbf{v}_{F}\cdot \mathbf{\nabla }\right) +E\right]
\eta \left( \mathbf{r}\right) +\Delta \left( \mathbf{r}\right) \chi \left(
\mathbf{r}\right)  &=&0,  \nonumber\\
\left[ i\hbar \left( \mathbf{v}_{F}\cdot \mathbf{\nabla }\right) -E\right]
\chi \left( \mathbf{r}\right) +\Delta \left( \mathbf{r}\right) \eta \left(
\mathbf{r}\right)  &=&0.  \label{Andreev equations}
\end{eqnarray}
It is assumed that the pairing gap function takes the steplike
form, $\Delta(x,y,z)=\Delta e^{i\phi}\theta(x)$ with $\phi$ being the SC phase.
In reality, it sags a little near the interface at the distance of
the mean free path or so. In the following, we neglect the above so-called
proximity effect since it gives only small corrections. As the
wave-vector components parallel to the interface, $k_y$, are
conserved for all possible processes, the problem is effectively
reduced to one-dimensional one.

The two ($\alpha$ and $\beta$) bands are assumed to be decoupled completely and thus one
can proceed to obtain various reflection coefficients for each
individual band. Outside of the interface ($x<0$), electrons and holes, if
existing, are free electrons and free holes and we denote their wave
functions by
\begin{equation}
\psi _{e}^{\pm }(x)=\left(
\begin{array}{c}
1 \\
0
\end{array}
\right) e^{\pm ik_+ x} ; \hspace{0.2in} \psi _{h}^{\pm }(x)=\left(
\begin{array}{c}
0 \\
1
\end{array}
\right) e^{\pm ik_{-}x} 
\end{equation}
with $k_y$ suppressed.
As is easily seen from the BdG equations, the wave vectors $k_{\pm }=k_{F}\sqrt{1\pm {E}/{\mu }}$.
Here $e$ ($h$) denotes for electron (hole) and $+$ ($-$) corresponds to
movement parallel (antiparallel) to the $x$-axis.

Inside the superconductor, one can solve the Andreev equation (\ref{Andreev
equations}) to obtain electron- and hole-like quasi-particle wave functions
\begin{eqnarray}
\Psi _{e}^{\pm }(x) &=&\left(
\begin{array}{c}
ue^{i\phi /2} \\
ve^{-i\phi /2}%
\end{array}%
\right) e^{\pm iq_{+}x},  \label{the vector in SC side} \\
\Psi _{h}^{\pm }(x) &=&\left(
\begin{array}{c}
ve^{i\phi /2} \\
ue^{-i\phi /2}%
\end{array}%
\right) e^{\pm iq_{-}x}.
\end{eqnarray}
Here the dispersion laws and expressions for $u$, $v$ also follow the BdG equations
together with the condition $u^2+v^2=1$. After some algebra, we obtain
\begin{eqnarray}
u\left( E\right)  &=&\sqrt{\frac{1+\sqrt{1-\Delta ^{2}/E^{2}}}{2}},
\label{coherent factors u} \\
v\left( E\right)  &=&\sqrt{\frac{1-\sqrt{1-\Delta ^{2}/E^{2}}}{2}},
\label{coherent factors v}
\\
q_{\pm } &=&k_{F}\sqrt{1\pm \frac{\sqrt{E^{2}-\Delta ^{2}}}{\mu }}.
\end{eqnarray}
For sub-gap excitations ($E<\Delta$), $u$, $v$, and $q$ will acquire imaginary parts.
Physically possible sub-gap solutions must decay into the bulk of the superconductor.
In the present case, it allows for $\Psi _{e}^{+}$ and $\Psi _{h}^{-}$ only
at $x\rightarrow\infty$.

To investigate the tunneling spectroscopy of a normal metal-superconductor point contact located
at $x=0$, we consider an electron that is incident on the boundary from the normal metal side $x<0$.
For elastic scattering,  $\psi _{e}^{+}$, $\psi _{e}^{-}$, and $\psi _{h}^{+}$ are 3 possible
quasi-particle states at the normal metal side with the same energy. Hence the quasi-particle wave function $\Psi_N$ at $x<0$ 
can be generally written as 
\begin{equation}
\Psi_N= \psi _{e}^{+}+r_{ee}\psi _{e}^{-}+r_{eh}\psi _{h}^{+}, \label{normal}
\end{equation}
where $r_{ee}$ and $r_{eh}$ characterize two possible reflection amplitudes with $r_{ee}$ describing
the normal reflection (reflected as electrons) and $r_{eh}$ describing the Andreev reflection (reflected as holes).
On the SC side
($x>0$), the quasi-particle wavefunction $\Psi_C$ is also a superposition of  electron-like ($\Psi_{e}$) and hole-like ($\Psi_{h}$) wave functions 
and can be generally written as  
\begin{eqnarray}
\Psi_{S} = t_{ee} \Psi_{e}^+ + t_{eh} \Psi_{h}^- , \label{SCside}
\end{eqnarray}
where $t_{ee}$ and $t_{eh}$ are transmission amplitude for electron-like and hole-like quasi particles. For multi-orbital superconductors,
elastic scatterings near boundary generally mixes quasi-particles of different bands at the same energy. 
For the model we adopt
for Fe-pnictide superconductors, there are 4 possible quasi-particle states in the SC side with the same energy: 
$\Psi^{+}_{e,\alpha}$, $
\Psi^{+}_{e,\beta}$, $\Psi^{-}_{h,\alpha}$ and $\Psi^{-}_{h,\beta}$. Hence the electron-like and hole-like wave 
functions can be generally written as 
\begin{eqnarray}
\Psi_{e}^+= A \Psi^{+}_{e,\alpha}+B
\Psi^{+}_{e,\beta} \nonumber \\
\Psi_{h}^-=C\Psi^{-}_{h,\alpha}+ D\Psi^{-}_{h,\beta}, \label{general}
\end{eqnarray}
where $A$, $B$, $C$, and $D$ are amplitudes for quasi-particles and quasi-holes in $\alpha$ and $\beta$ bands.
Combining Eqs.(\ref{SCside}) and (\ref{general}), it is clear that coefficients $A$ and $D$ can be absorbed into
$t_{ee}$ and $t_{eh}$ and the electron-like and hole-like wave 
functions can be recast into the following forms
\begin{eqnarray}
\Psi_{e}^+= \Psi^{+}_{e,\alpha}+\gamma
\Psi^{+}_{e,\beta} \label{general1} \\
\Psi_{h}^-=\gamma' \Psi^{-}_{h,\alpha}+ \Psi^{-}_{h,\beta}. \label{general2}
\end{eqnarray}
Here $\gamma'$ ($=C/D$) and $\gamma$ ($=A/B$) are two coefficients that characterize coupling
of $\alpha$ and $\beta$ bands. For bulk superconductors without boundaries, 
quasi-particles in $\alpha$ and $\beta$ bands are not mixed.
Hence $\gamma$ and $\gamma's$ are independent parameters in this case.  
However, the boundary introduces definite coupling between $\alpha$ and $\beta$ bands. Therefore,
$\gamma$ and $\gamma's$ are no longer independent parameters. In the simplest
assumption\cite{PhysRevLett.103.077003}, one sets $1/\gamma' = \gamma \equiv \alpha_0$. In this case,
quasi-particles and quasi-holes have the same coupling ratio in $\alpha$ and $\beta$ bands. As indicated
in the introduction, the resulting Andreev bound states generally appear at finite
energies and do not explain the observed ZBCP in PCARS. 

In order to explain the observed ZBCP in PCARS, we first note that the boundary scatterings
for quasi-particles and quasi-holes of the same band are generally different due to that there is no particle-hole symmetry in the normal metal side. The asymmetry between quasi-particles and quasi-holes is exhibited in
Eqs.({\ref{general1}) and (\ref{general2}) as the factor $\gamma'$ for the $\alpha$ band and $\gamma$ for the $\beta$ band.
Note that there is a tendency to preserve an overall particle-hole symmetry in the superconducting side as
reflected in Eqs.({\ref{general1}) and (\ref{general2}), where the asymmetry for the $\beta$ band is reversed and is the ratio
of the amplitude of quasi-particle to that of the quasi-hole.  Physically, Eq.(\ref{general1}) implies that a quasi-particle in $\alpha$ band can scatter into $\beta$ band. Since when a quasi-particle transmits from $\alpha$ band into $\beta$ band, a hole is left in $\alpha$ band. Hence a quasi-particle tunneling from $\alpha$ band into $\beta$ band is equivalent to a quasi-hole tunneling from $\beta$ band into $\alpha$ band, which, when combined with Eq.(\ref{general2}), implies $\gamma'=\gamma$. Therefore, we consider a class of parameters in which
\begin{equation}
\gamma' = \gamma
\end{equation}
is obeyed. In this class of parameters, a large probability amplitude for a quasi-particle tunneling from $\alpha$-band to $\beta$-band will result in a large probability amplitude for a quasi-hole tunneling from $\beta$-band to $\alpha$-band.  
As a result, unlike the phenomenological ratio of probability amplitudes
$\alpha_0$\cite{PhysRevLett.103.077003}, asymmetry in the tunneling probability amplitude of $\alpha$ and $\beta$ would induce asymmetry between particles and holes in the same orbital.   
We shall show in subsection \ref{Comparing with the experimental data} that the introduction of asymmetric factor $\gamma$ can naturally explain experimental data.

In addition to the above consideration on mixing of $\alpha$ and $\beta$ bands, the potential scattering near the boundary can be generally described by a delta-function barrier potential,
$H\delta\left( x\right)$.  Matching the wave functions and their derivatives at the
interface ($x=0$)\cite{PhysRevB.25.4515}
\begin{eqnarray}
\Psi_{N}\left( x\right)  |_{x=0^{-}}  &  =\Psi_{S}\left(  x\right)
|_{x=0^{+}},\label{eq:boundary conditions} \nonumber \\
\frac{2mH}{\hbar^{2}}\Psi_{S}\left(  x\right)  |_{x=0^{+}}  & =\frac
{d\Psi_{S}\left( x\right)
}{dx}|_{x=0^{+}}-\frac{d\Psi_{N}\left(x\right)
}{dx}|_{x=0^{-}}, \label{match}
\end{eqnarray}
yields complete solutions of all scattering amplitudes. Note that the boundary conditions imposed in Eq.(\ref{match}) actually consist of four equations and represent boundray conditions directly on electron wave functions.
After some tedious but straightforward
derivations, we find that  when $\phi_\alpha-\phi_\beta=\pi$ or $0$, reflection coefficients are given by
\begin{eqnarray}
\Lambda r_{eh} &=&\left( \gamma u_{\alpha}+\epsilon u_{\beta}\right) \left(
v_{\alpha}+\epsilon \gamma v_{\beta}\right) , \nonumber \\
\Lambda r_{ee} &=&\left( iZ+Z^{2}\right) \left[ \left( \epsilon v_{\alpha}+\gamma
v_{\beta}\right) \left( \gamma v_{\alpha}+v_{\beta}\right) -\right.  \nonumber \\
&&\left. \left( u_{\alpha}+\gamma u_{\beta}\right) \left( \epsilon \gamma
u_{\alpha}+u_{\beta}\right) \right] . \label{reflectionC}
\end{eqnarray}
Here $Z=2mH/\hbar^2 k_F$, $\Lambda =( 1+Z^{2}) ( \epsilon \gamma u_{\alpha}+u_{\beta})
( u_{\alpha}+\gamma u_{\beta}) -Z^{2}( \gamma v_{\alpha}+v_{\beta})
( \epsilon v_{\alpha}+\gamma v_{\beta})$, and when $\phi_\alpha-\phi_\beta=0$, $\epsilon=1$; otherwise,  $\epsilon=-1$. The tunneling conductance is correspondingly given by the two
solved coefficients $r_{ee}$ and $r_{eh}$ as 
\begin{equation}
\sigma_S=1+|r_{eh}|^2-|r_{ee}|^2. \label{tunnelingconductance}
\end{equation}

\section{Results and Discussions} \label{Results and Discussions}

\subsection{Numerical simulations}\label{Numerical simulations}

\begin{figure}[ptb]
\includegraphics[width=9cm ]{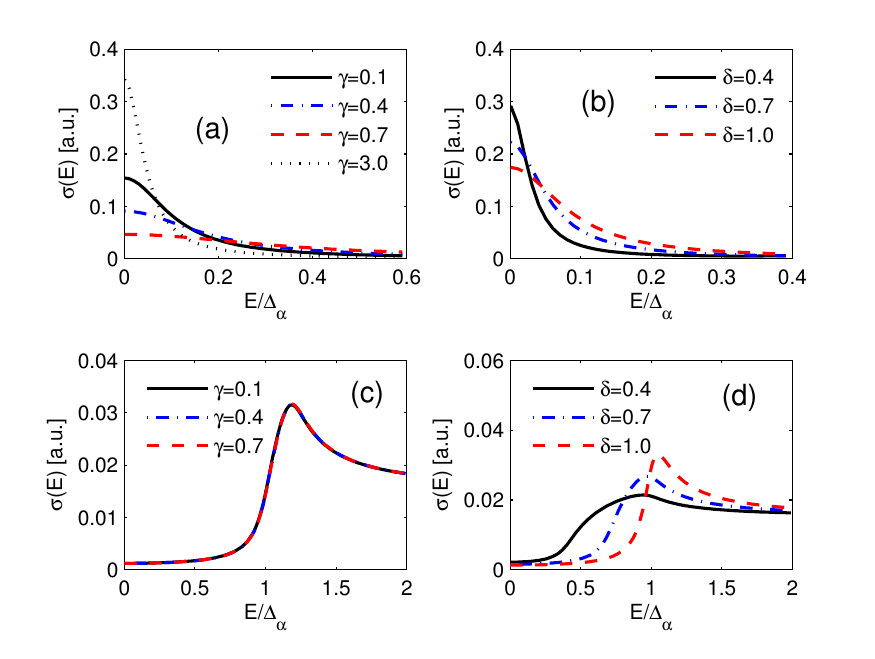}
\caption{(Color online) Panel (a) and (b): Tunneling conductance for the pairing symmetry $s^{\pm}$. Panel (c) and (d): Tunneling conductance for the pairing symmetry $s^{++}$. Here $\delta=\Delta_{\beta}/\Delta_{\alpha}$ and $k_y$ of the incident electron is zero. Parameters in Panel (a) and (c) are  $\delta =1.2$ and in Panel (b) and (d) are $\gamma=0.1$. Except for (c) where we set $Z=1$, $Z$ is taken to be $8$ in other cases. Note that a scattering broadening $\Gamma=0.08$ is used to prevent divergence in all cases.}  \label{fig2}
\end{figure}

Fig. \ref{fig2} shows the zero-temperature conductance for
SC pairing $s^{\pm}$ of two bands when $k_y$ of the incident electron is zero. Sharp ZBCP emerges in (a) and (b) when the
mixing coefficient $\gamma$  is take at very small or very large values. The sharp ZBCP can be attributed to
the existence of zero-energy Andreev bound states. Indeed, Eq.({\ref{reflectionC}) implies
that a bound state exists when 
\begin{equation}
\Lambda = 0.
\end{equation}
For large Z,  since $\Lambda \sim Z^2 \{(\epsilon \gamma u_{\alpha}+u_{\beta})
( u_{\alpha}+\gamma u_{\beta}) -( \gamma v_{\alpha}+v_{\beta})
( \epsilon v_{\alpha}+\gamma v_{\beta}) \}$, we find that in the limit of large $\gamma$, $\Lambda =-\gamma(u_{\alpha}u_{\beta}-v_{\alpha}v_{\beta})$, while for $\gamma \rightarrow 0$, $\Lambda =(u_{\alpha}u_{\beta}-v_{\alpha}v_{\beta})$. 
Because $u_{\alpha}u_{\beta}-v_{\alpha}v_{\beta}$ vanishes when $E=0$, both $\gamma \rightarrow \infty$ and $\gamma \rightarrow 0$ support zero energy solution. Hence zero-energy bound state exists in these limits. When $\gamma$ is not too large or non-vanishing, the zero-energy bound state is partially destructed.
For comparison, we also present results for the pairing symmetry $s^{++}$, shown in Figs. \ref{fig2} (c) and (d). Clearly, the zero-energy bound states are absent. The tunneling conductance exhibits one quasi-particle peak which indicates that the conductance is not a simple
sum over two individual bands. In particular, the numerical solutions indicate that the position of quasi-particle peak is insensitive to the mixing constant $\gamma$ and is sensitive to the ratio of two gap amplitudes, $\delta$.

\subsection{Quantization condition of midgap states and zero energy bound state} \label{Zero energy bound state}

In order to understand the arising of ZBCP in limits of $\gamma \rightarrow 0$ and $\gamma \rightarrow \infty$, we further 
examine the zero energy bound state by investigating multiple scatterings of quasi-particles. The examination will
generally yield the quantization condition for existence of midgap states and the condition for the
existence of zero energy bound state.

For this purpose, we consider quasi-particles in a thin normal metal layer of width $L$
attached to a Fe-pnictide superconductor as shown in Fig. ({\ref{fig3}).
In the limit of $L \rightarrow 0$, as first shown by Hu\cite{Hu1526}, 
semi-classical quantization rule is sufficient to determine midgap states.
For this purpose, we first consider quasi-particles with energy $E$ off from a superconductor
with a single band at $x>0$ with gap magnitude being $\Delta$. 
Following the same procedure outlined in Sec. \ref{Model and Method}, 
matching the wave functions and their derivatives at the
interface ($x=0$),  one obtains the coefficients
$r_{eh}$ and $r_{ee}$. These coefficients can be greatly simplified under the so-called
Andreev approximation. In the lowest non-vanishing order
in $\max(\Delta,E)/\mu$, one can take $k_\pm\approx q_\pm\approx k_F$ and
\begin{equation}
k_{+}-q_{-}\approx q_{+}-k_{-}\approx k_{F}\frac{u}{v}\frac{\Delta }{2\mu }.
\end{equation}
Therefore, we find that the Andreev reflection coefficients can be approximated as\cite{PhysRevB.25.4515,Zagoskin}
\begin{equation}
r_{eh(he)}=e^{\mp i\phi }\times \left\{
\begin{array}{c}
e^{-i\cos^{-1} \frac{E}{\Delta }},\rm{~for~~} E\leq \Delta,  \\
e^{-\rm{cosh^{-1}} \frac{E}{\Delta }},\rm{~for~~} E>\Delta.
\end{array}
\right.   \label{Andreev reflection amplitude}
\end{equation}
For the sub-gap (or midgap) state, $E\leq \Delta$, we thus have for
the total Andreev reflection $|r_{eh(he)}(E)|^2 = 1$ and
consequently the normal reflection can be safely neglected.

When a subgap-energy electron ($E<\Delta$) enters the interface, an Andreev hole will reflect from the interface due to the electron-hole coupling through the pairing potential $\Delta$. Since there are two bands at $x>0$, an electron can
be reflected either by the $\alpha$ band or by the $\beta$ band. Therefore, it is convenient to keep track on band origin of quasi-particles in the metal layer by writing the wavefunction $\psi (x,E)$ of quasi-particle for $-L<x<0$ as
\begin{eqnarray}
\psi (x,E) = && a\psi _{\alpha e}^{+}\left( x,E\right)
+ b \psi _{\alpha h}^{+}\left( x,E\right) + c \psi _{\beta e}^{-}\left( x,E\right) +   
d\psi _{\beta h}^{-}\left( x,E\right) \nonumber \\
&&  + a'\psi _{\beta e}^{+}\left( x,E\right) +   
b' \psi _{\beta h}^{+}\left( x,E\right) + c'\psi _{\alpha e}^{-}\left( x,E\right)
+ d'\psi _{\alpha h}^{-}\left( x,E\right). \label{wavefunction}
\end{eqnarray}
Here $a$, $b$, $c$, $d$, $a'$, $b'$, $c'$, and $d'$ are coefficients to
be determined by boundary conditions and we have set $k_y=0$ for sake of illustration. As illustrated in Fig.\ref{fig3}, within the metal layer, there are several possible close cycles, known as Saint-James cycles
\cite{RevModPhys.77.109}, that quasi-particles may form during their scatterings at $x=0$ and $x=-L$. For instance, an $\alpha$ electron particle may be reflected as an $\alpha$ hole at $x=0$ and the $\alpha$ hole gets specular reflection at $x=-L$. The specularly reflected $\alpha$ hole will scatter back as an $\alpha$ electron at $x=0$ and form an intra-band close cycle. Similar intra-band close cycles also exist for the $\beta$ band. Due to {\em non-conservation of momentum} perpendicular to the interface at $x=0$ during scattering, there are also inter-band Saint-James cycles.
As explained, the asymmetry between particles and holes due to scatterings is taken into account by $\gamma$ in Eqs.(\ref{general1}) and (\ref{general2}). In the limit of $\gamma \rightarrow 0$, $\Psi^{+}_{e,\alpha}$ and $\Psi^{-}_{h,\beta}$ (equivalently, $\Psi^{+}_{h,\alpha}$ and $\Psi^{-}_{e,\beta}$) dominate in the SC side. Since quasi-particle wavefunctions are continuous across the junction $x=0$, solutions with vanishing $a'$, $b'$, $c'$, and $d'$ correspond
to the limit of $\gamma \rightarrow 0$. Similarly, solutions with vanishing $a$, $b$, $c$, and $d$ correspond
to the limit of $\gamma \rightarrow \infty$. 

\begin{figure}[ptb]
\begin{center}
\includegraphics[
width=8cm ]{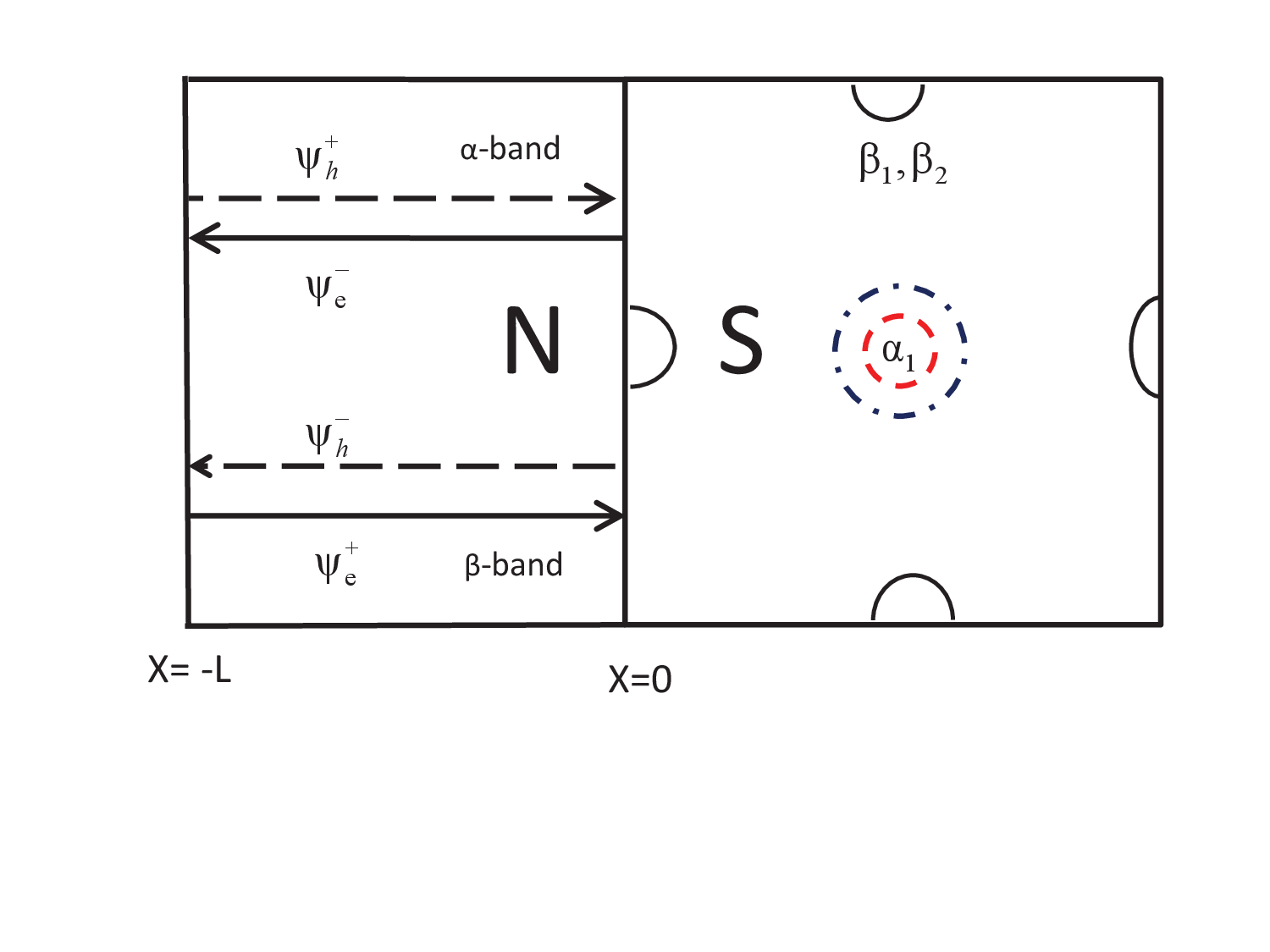}\vspace{-2.0cm}
\caption{(Color online) Schematic representation of
an Andreev-Saint-James cycle for a two-band $s^{\pm}$-wave superconductor
coated with a normal-metal layer, the interface being
oriented perpendicular to the [100] direction.}
\label{fig3}
\end{center}
\end{figure}
Quantitatively, Eq.(\ref{normal}) implies that in each scattering, $\psi_e^+$ has to match $r_{eh} \psi_h^+$
and  $\psi_h^-$ has to match $r_{he} \psi_e^-$. Equivalently, 
$\psi_e^-$ has to match $r_{he} \psi_h^-$
and  $\psi_h^+$ has to match $r_{eh} \psi_e^+$. 
Together with the hard wall boundary condition, $\psi(-L)=0$, applied to
particles and holes separately, we find that the inter-band cycle in the limit $\gamma \rightarrow 0$ implies the following relations
\begin{eqnarray}
a \psi _{\alpha e}^{+}\left( 0,E\right)  &=&r_{eh} (\alpha) b \psi _{\alpha h}^{+}\left(
0,E\right), \nonumber \\
d\psi _{\beta h}^{-}\left( 0,E\right) &=& r_{he} (\beta) c \psi _{\beta
e}^{-}  \left( 0,E\right), \nonumber \\
c\psi _{\beta e}^{-}\left( -L,E\right) &=& -a\psi _{\alpha
e}^{+}\left(-L,E\right) , \nonumber \\ 
b\psi _{\alpha h}^{+}\left( -L,E\right) &=& -d \psi _{\beta h}^{-}\left( -L,E\right),  \label{Saint-James cycle}
\end{eqnarray}
where the last two equations result from the specular reflections at $x=-L$. Clearly, by taking $L \rightarrow 0$, Eq.~(\ref{Saint-James cycle}) implies  $r_{eh}(\alpha) r_{he}(\beta) =1$. Similarly, for $\gamma \rightarrow \infty$, one replaces $a$, $b$, $c$, and $d$ by $a'$, $b'$, $c'$, and $d'$. Furthermore, $\alpha$ and $\beta$ in (\ref{Saint-James cycle}) get exchanged. Hence we
get a reversed cycle and obtain $r_{eh}(\beta) r_{he}(\alpha)=1$. By
using Eq. (\ref{Andreev reflection amplitude}), we conclude that the
midgap energy in a semi-infinite Fe-pnictide superconductor that occupy $x>0$  must satisfy
\begin{equation}
\cos^{-1} \frac{E_{n}}{\Delta _{\alpha }}+\cos^{-1} \frac{E_{n}}{\Delta _{\beta }
} =\pm \left( \phi _{\alpha }-\phi _{\beta }\right) +2n \pi,  \label{Andreev levels}
\end{equation}
where $n=0$, $\pm1$, $\pm 2$, $\cdots$.
Eq.~(\ref{Andreev levels}) represents one of the major results in
this paper. If one identifies the scattering phase across the
interface as the generalized momentum,
$p_i=\pm\phi_i+\cos^{-1}\frac{E}{\Delta_i}$, Eq.~(\ref{Andreev
levels}) can be rewritten as $\sum_{i=(\alpha, \beta)} p_i=2n\pi$,
which is in consistence with the semiclassical quantization
condition. When $\Delta\phi=\phi_\alpha-\phi_\beta=\pm\pi$
$E_n< \rm{min(}\Delta_\alpha$,$\Delta_\beta$), Eq.~(\ref{Andreev levels}) only
supports the zero-energy solution ($E_n=0$). It indicates that
zero-energy surface bound states can exist in the semiclassical
approximation even in
the limit of a zero-thickness normal slab.
The zero-energy states
are formed in the normal side and extend into the superconducting side over a coherence
length which is similar the case of $d$-wave superconductor studied by Hu\cite{Hu1526}.
Moreover the zero-energy surface bound state is only sensitive
to the phase difference of the two-band pairing and has nothing
to do with the paring amplitudes. It
should be noted that no Saint-James cycle exists when
$\Delta_\alpha<E_n<\Delta_\beta$ or $\Delta_\alpha<\Delta_\beta<E_n$
because according to Eq.~(\ref{Andreev reflection amplitude}), the
currents will decay exponentially. 

\subsection{Comparison with the experimental data} \label{Comparing with the experimental data}

To consider how midgap
states are related to real PCARS measurements, we first note that for point-contact junctions,
even though the interface of the junction is not an infinitely flat plane, the contact region is flat and is often large in comparison to the atomic scale and the BTK theory is applicable\cite{Greene}. However, in contrast to planar junctions,
point-contact junctions are often plagued with uncertainty in exact orientation of the junction interface. Therefore, it is crucial to include orientational dependence in the tunneling spectroscopy. For this purpose, in the following we extend the
analysis to include $k_y \neq 0$. 

We first note that if the pairing
gaps between different FSs are extended $s$-wave with sign reversal,
there will be four kinds of zero-energy surface bound states (or
zero-energy Saint-James cycles) corresponding to various
combinations of interband FSs: $\alpha_1-\beta_1$,
$\alpha_1-\beta_2$, $\alpha_2-\beta_1$, and $\alpha_2-\beta_2$. As indicated in Fig. \ref{fig3}, zero-energy Saint-James cycles is sensitive to the direction. It can only exist when only when quasi-particles can scatter
across $\alpha$ and $\beta$ bands. The window for such scatterings can be estimated as
the spanning angle of $\beta$ band with respect to the $\mathbf{k}=0$ point, which limits the orientations of
interfaces to fall into a small window around $\pm 6^{\circ}$ near the
[100] direction. In contrast to other pairing symmetry such as $d$-wave which also supports
zero-energy bound states but with a large window for
observing these states\cite{PhysRevB.67.024503}.

In addition to the above four zero-energy surface bound states, there
also exist nonzero-energy surface bound states (or nonzero-energy
Saint-James cycles) corresponding to various combinations of
intraband FSs: $\alpha_1-\alpha_1$, $\alpha_1-\alpha_2$, $\alpha_2-\alpha_2$,
$\beta_1-\beta_1$, $\beta_1-\beta_2$ and $\beta_2-\beta_2$. This is
the key why two coherent peaks are often observed in the PCARS measurements.
It has also been pointed out that nonzero ABS can exist in the $s^\pm$-wave
superconductor of multi-gap nature due to interference effects\cite{PhysRevB.79.174529, PhysRevLett.103.077003}
and inter-band quasi-particle scattering\cite{PhysRevB.79.174526}. These effects may determine
whether an electronic trajectory crosses two bands or only one band. It may also contribute to sub-gap peaks in the angular dependence of conductance spectra for $s^{\pm}$-wave Fe-pnictide superconductors.

To realize the above analysis in real PCARS, we calculate the conductance in $\gamma\gg1$ case. The results in  $\gamma\ll1$ case will remain unchanged. In these limits, results of the tunneling spectroscopy can be simplified. 
On the SC side ($x>0$), for a given $k_y$ and $E$, since $k_y$ is
conserved, available quasiparticles depends on the magnitude of $k_y$ .
For $k_y \sim 0$, i.e., near the [100] direction as shown in Fig.~\ref{fig3},
 the transited QPs can be either on the $\alpha$ band or $\beta$ band, experiencing different
pairing potentials for $s^{\pm}$-wave Fe-pnictide superconductors.
This enables the completion of zero-energy Saint-James cycle.
We shall set the electron-like QP wavefunction $\Psi^+_e$ in Eq.(\ref{SCside}) to be in the quasi-particle wavefunction of the $\alpha$ band with pairing potential
$\Delta_\alpha e^{i\phi_\alpha}$. Similarly, the hole-like QP wave
function $\Psi^+_h$ in Eq.(\ref{SCside}) is obtained by using pairing potential $\Delta_\beta
e^{i\phi_\beta}$. Assuming electrons are incident with an angle $\theta$ with respect to the [100] direction,
one can set $k_y = k_F \sin \theta$.
By matching the wave
functions and their derivatives at the interface $x=0$ according to Eq.(\ref{match}), the normalized tunneling conductance $\tilde{\sigma}_{S}$ is obtained via Eq.(\ref{tunnelingconductance}). We obtain
\begin{eqnarray}
\tilde{\sigma}_{S}\left( E\right)=
\frac{16\left( 1+\cos ^{2}\theta \left\vert \Gamma _{\alpha }\right\vert
^{2}\right) \cos ^{4}\theta +4Z^{2}\left( 1-\left\vert \Gamma _{\alpha
}\Gamma _{\beta }\right\vert ^{2}\right) \cos ^{2}\theta }{\left\vert 4\cos
^{2}\theta +Z^{2}-Z^{2}\left\vert \Gamma _{\alpha }\Gamma _{\beta
}\right\vert \exp \left[ i\left( \phi _{\alpha }-\phi _{\beta }\right) %
\right] \right\vert ^{2}}.
\label{the tunneling conductance}
\end{eqnarray}
Here $\Gamma$'s are defined as $\Gamma _{\alpha ,\beta }=E / \vert \Delta _{\alpha ,\beta
} \vert-\sqrt{ (E / \vert \Delta _{\alpha ,\beta
} \vert ) ^{2}-1}$. 
Similar derivation of Eq.~(\ref{the tunneling conductance}) 
can be found in Ref. \cite{Tanaka3451} where a single-band system
was studied.

\begin{figure}[ptb]
\includegraphics[
width=8cm ]{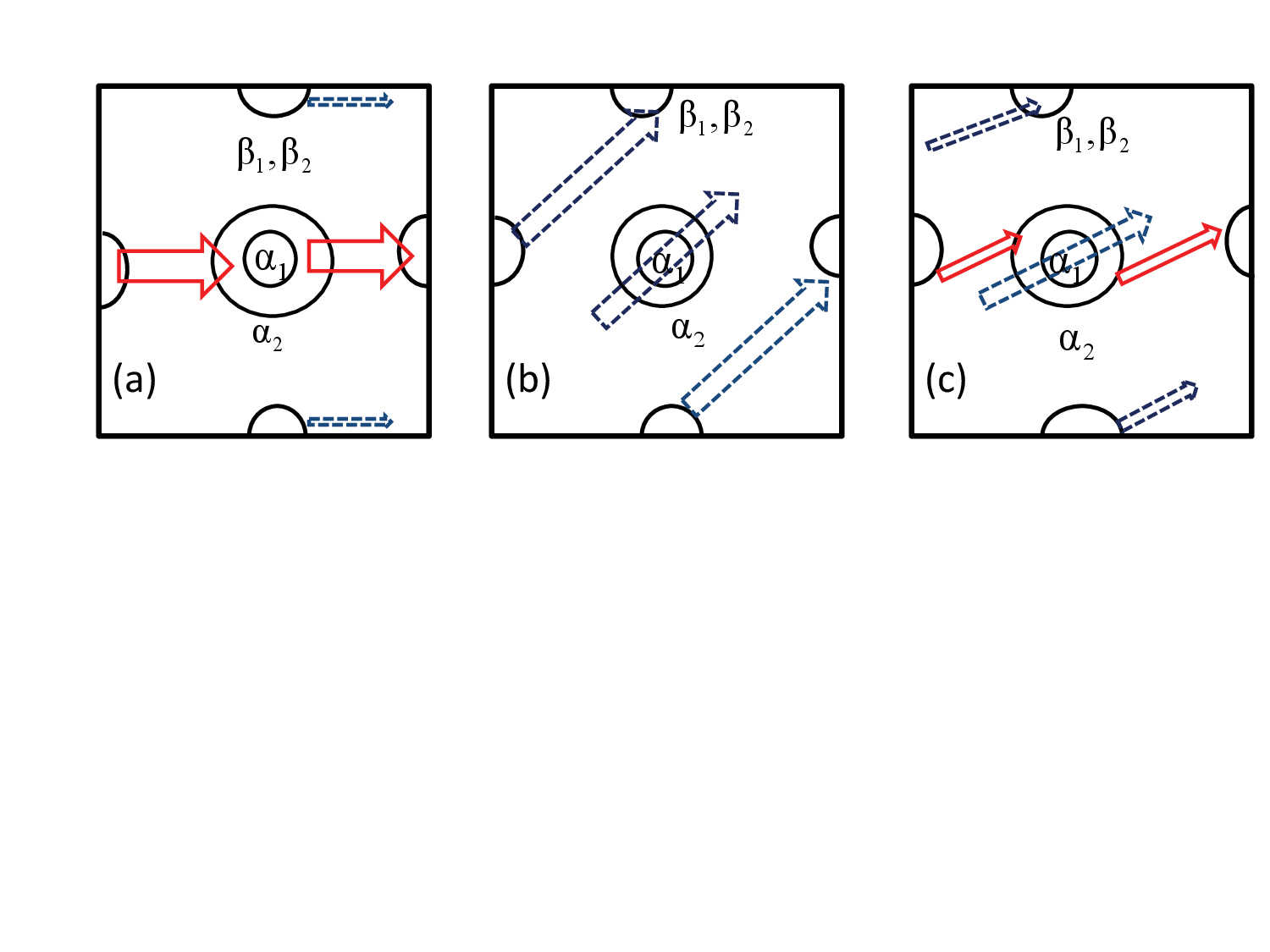}\vspace{-3.0cm}\\
\hspace{-0.5cm}
\includegraphics[
width=8cm ]{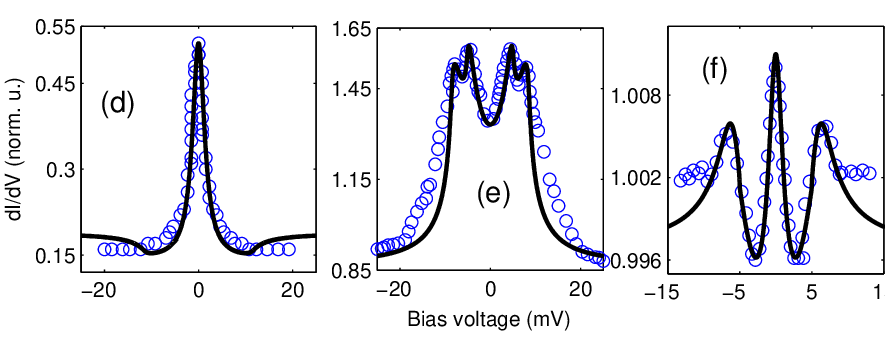}
\caption{(Color online) Panel (a)--(c): schematic plots of different-direction electron tunneling
on iron-pnictide SC interfaces. The arrows denote the electron tunneling directions and how
different bands are coupled. The arrows with red sold lines show the particular
directions where zero-energy Saint-James cycle can form.
The weights of the arrows denote their possible contributions to the differential conductance.
Panel (d)--(f): fitting to the I-V data of (d) SmFeAsO$_{0.9}$F$_{0.1}$ sample ($T_c=51.5$K)
(taken from Ref. [\cite{0295-5075-83-5-57004}]); (e) TbFeAsO$_{0.9}$F$_{0.1}$ sample ($T_c=50$K)
(taken from Ref. [\cite{1367-2630-11-2-025015}]); and (f) SmFeAsO$_{0.9}$F$_{0.1}$ sample ($T_c=51.5$K)
(taken from Ref. [\cite{0953-2048-22-1-015018}]).}
\label{fig4}
\end{figure}

Eq. (\ref{the tunneling conductance}) is valid only when  $\theta \sim 0$ (and thus $k_y \sim 0$).
In this case, as shown in Fig.~\ref{fig4}(a), due to
non-conservation of $k_x$, the dominant contribution to
the current comes from QPs from both $\alpha$ and $\beta$ bands. The low-energy
differential conductance will thus be dominated by the zero-energy
bound states that arise from the inter-band QP transitions although the
nonzero-energy bound states may also have some effects on it. Thus
the ZBCP found in PCARS is due to the zero-energy bound states that
exist in the interface when electrons tunnel normal to the
interface along the [100] direction. This is verified in
Fig.~\ref{fig4}(d), where we have simulated the PCARS data presented
in Ref. \cite{0953-2048-22-1-015018} by assuming that all electrons
are tunneling across the interface along the [100]-direction.
Eq.~(\ref{the tunneling conductance}) is used to obtain the
best fitting with $\Delta_\alpha=8$ and $\Delta_\beta=8$
under the proposed sign-reversal $s^\pm$-wave pairing.
Here the integration over angles is not necessary for the small window of angles.
The normalized barrier height is taken to be $Z=5$ and the scattering
broadening $\Gamma=0.5$.

It should be emphasized that ZBCP is
more sensitive to the phase difference than the gap amplitude and
when a more realistic FS together with a more realistic extended
$s$-wave gap are used, a much better fitting, especially in the
nonzero-energy region, will be obtained. The ZBCP found here can
only exist when range of $k_y$ covers the $\beta$ band. Using
experimental data\cite{0295-5075-83-4-47001, PhysRevB.83.020501},
this corresponds to a small window of $\pm 6^{\circ}$ for interfaces
around [100] direction.

We next consider the case of the [110] direction in Fig.~\ref{fig4}(b)
for which the interface is off the window of $\pm 6^{\circ}$ about [100].
In this case, the conservation of $k_y$ removes QPs in the $\beta$ band.
Therefore, the crossing term $\Gamma_{\alpha} \Gamma_{\beta}$ in
Eq. (\ref{the tunneling conductance}) is absent. Hence the denominator
in Eq. (\ref{the tunneling conductance}) is always finite and it
results in in the absence of
the ZBCP. In this case, however, non-zero Saint-James cycles of two separate bands
are important. This is similar to the point-contact conductance
spectra of MgB$_2$, which is also a multi-band superconductor and
the tunneling spectrum is usually fitted by summation of two single-band
tunneling probabilities\cite{PhysRevLett.89.247004, PhysRevB.65.180517}.
In Fig.~\ref{fig4}(e), we use the two-gap $s^\pm$-wave
model to fit the PCARS data reported in Ref. \cite{1367-2630-11-2-025015}
by using Eq.(\ref{the tunneling conductance}). The best fitting was obtained
with the two gap amplitudes set to $\Delta_\alpha=8.5$ and
$\Delta_\beta=5$. The broadening is taken to be $\Gamma=0.5$.

In Fig.~\ref{fig4}(c), we consider the case when the orientation of
the interface is between [100] and [110] but falls into the window
of $\pm 6^{\circ}$ about [100]. In such case, both nonzero- and
zero-energy bound states are equally important to the differential
conductances and ZBCP is often observed due to zero-energy midgap
state. Moreover, when electrons tunnel into the superconductor
across the interface, QPs from different bands are coupled. Due to
different gap amplitudes on different FSs and finite QP life time,
one effective gap amplitude can be generally observed in experiment.
In Fig.~\ref{fig4}(f), we have used one gap amplitude sign reversal
[as done in Fig.~\ref{fig4}(a)] to fit the PCARS data
\cite{0953-2048-22-1-015018}. For the best fitting, the effective
gap amplitude is taken to be $\overline{\Delta}=6.5$ with broadening
being $\Gamma=1.5$, while the sign-change gap amplitude is taken to
be $\Delta=5$ with the broadening being $\Gamma=0.01$.

\section{summary} \label{summary}

In summary,  we have derived a phenomenological model to account for 
the observed tunneling spectroscopy of Fe-pnictide superconductors by taking into the consideration of asymmetric interface scattering between particle and holes. Signatures of anti-phase
$s^{\pm}$-wave pairing in Fe-pnictide superconductors are shown to exhibited as 
zero-energy surface bound states. In contrast to other
pairing symmetries such as $d$-wave that also supports zero-energy bound
states but of a large window of interface orientation for observing these
states, for $s^{\pm}$-wave, zero-energy
bound states can exist only when the orientations of interfaces fall
into a small window around $\pm 6^{\circ}$ near the [100] direction.
Off the [100] direction, the zero-bias peak disappears and is
replaced by two coherent peaks, due to directional dependence of QPs
interplaying between different bands. Our results give a unified
explanation to the Point-contact Andreev-reflection spectroscopy (PCARS)
data in various directions and indicate strongly that current PCARS
favor the scenario of $s^{\pm}$-wave pairing for multi-band Fe-pnictide superconductors .

Finally it is worth noting that PCARS results are very sensitive to the variation
and orientation of the micro-crystals. It is indeed a local probe.
On the other hand, zero-energy Saint-James cycles is shown to exist only
in a small window around $\pm 6^{\circ}$ near the [100] direction.
In our modeling, PCARS can only detect the phase difference
between the two bands with the same orientation. Moreover,
$d_{x^2-y^2}$-wave symmetry cannot be ruled out if
the phase difference between the two bands is $\pi$\cite{0953-2048-22-1-015018}.
The effects of interface disorder may also be important, which will lead to
scattering between momentum states along the interface. The PCARS should be
more complex than one expects. Only when high-quality single crystal is available
and more experiments are performed, one can have a more concrete picture on the
pairing symmetry of iron-based superconductors.

\ack
This work was supported by National Science Council of Taiwan (Grant
Nos. 100-2811-M-007-015 and 100-2112-M-007-011-MY3), Hebei Provincial
Natural Science Foundation of China (Grant No. A2010001116), and the
National Natural Science Foundation of China (Grant No. 10974169). C. S. Liu would like to thank the hospitality of Institute of Applied Physics and Computational Mathematics at Beijingt where some of the work was carried out during his visit. We also acknowledge the support from the National Center for
Theoretical Sciences, Taiwan.

\end{document}